\documentclass{llncs}
\usepackage{amsfonts}
\usepackage{amssymb}
\usepackage{amsmath}

\usepackage{txfonts}
\usepackage{graphicx}
\usepackage{subfigure}
\usepackage{slashbox}
\usepackage{url}

\usepackage{bm}

\usepackage{subfigure}

\newcommand{\MB}{\[\begin{array}{lllll}}
\newcommand{\ME}{\end{array}\]}
\newcommand{\oomit}[1]{}

\newcommand{\noop}[1]{}

\newcommand{\comment}[1]{}
\newcommand{\bigonly}[1]{}

\begin{document}
\pagestyle{empty}


\title{Online Verification of Control Parameter Calculations in Communication Based Train Control System}


\author{Lei Bu, Xin Chen, Linzhang Wang and Xuandong Li} \institute{State Key Laboratory
for Novel Software Technology, Nanjing University\\
Department of Computer Science and Technology, Nanjing University
\\Nanjing, Jiangsu, P.R.China 210093\\ \email
{\{bulei$|$chenxin$|$lzwang$|$lxd\}@nju.edu.cn }}


\maketitle

\begin{abstract}

Communication Based Train Control (CBTC) system is the
state-of-the-art train control system. 
In a CBTC system, to guarantee the safety of train operation, trains
communicate with each other intensively and adjust their control
modes autonomously by computing critical control parameters, e.g.
velocity range, according to the information they get. As the
correctness of the control parameters generated are critical to the
safety of the system, a method to verify these parameters is a
strong desire in the area of train control system.

In this paper, we use our experience learned during verifying a CBTC
system to present our ideas of how to model and verify the control
parameter calculations in a CBTC system efficiently.
\begin{itemize}
\item As the behavior of the system is highly nondeterministic, it is difficult to build and verify the complete behavior space model of the system offline in advance. Thus, we propose to model the system according to the ongoing behavior model induced by the control parameters.
\item As the parameters are generated online and updated very quickly, say every 500 milliseconds in the case we met, the verification result will be meaningless if it is given beyond the time bound, since by that time the model will be changed already. Thus, we propose a method to verify the existence of certain dangerous scenarios in the model online quickly.
\end{itemize}


To demonstrate the feasibility of these proposed approaches, we
present the composed linear hybrid automata with readable shared
variables as a modeling language to model the control parameters
calculation and give a path-oriented reachability analysis technique
for the scenario-based verification of this model. We demonstrate
the model built for the CBTC system, and show the performance of our
technique in fast online verification. Last but not least, as CBTC
system is a typical CPS system, we also give a short discussion of
the potential directions for CPS verification in this paper.

\end{abstract}

\section{Introduction}

Nowadays, as communication has been embedded deeply into our daily life, computation has evolved from locating in one single standalone
device to the collaboration of networks of equipments. In such a manner, more and more systems work in open environments, receive
signals and stimuli from sensors, actuators and networks, then calculate their control modes and parameters accordingly. The newly
generated control modes and parameters will control the behavior of the system itself and the behavior of other components in the network
as well dynamically. These systems have a tight integration of information systems and physical devices, which are named as Cyber-Physical Systems (CPS)\cite{lee2}. 

By combining communication, computation and control (3C), in a CPS
system physical devices can have more knowledge of the environment
they are working in and the real-time status of the other elements
which they are collaborating with. Thus, devices can autonomously
generate more accurate instructions and gain advantages like safety,
reliability and efficiency.


Public transportation is a typical area where CPS systems are
emerging and playing more and more important roles. 
CBTC is the state-of-the-art technique in the train control area and
fundamental for the building and controlling of high speed railway
systems. During trains running on railways, the radio block center
(RBC) will collect the position of each train periodically and
compute the movement authority (MA), which is the distance that the
train is authorized to go, for each train. 
Then the onboard train controller will compute the feasible velocity
range by taking account of the movement authority and the current
running parameters of the train, e.g., current position, velocity
and etc.  These are typical procedures of a CBTC system, which is
clearly a
Cyber-Physical System. 
One of the most important questions concern the design engineers of
CBTC system is whether the parameters generated by the control
functions used in the system are correct, e.g. trains will not
collide with each other during operation.

In general, if we can build a model for the control parameter
calculations and verify it, we can answer the correctness of
parameters. Currently, most of the verification works consist of the
following two steps: First, build the complete static formal model
of the system. Second, verify the correctness of the model under the
given property offline using techniques like model
checking\cite{mc}. For CBTC systems, as the input of the control
functions, e.g. current velocity, position, movement authority and
etc., are generated and collected online, it is hard to predict the
complete behavior space of the system under verification.  Thus, it
is difficult to build and verify a complete static model of the
system's behavior  offline in advance. To overcome this problem, we
discuss our opinions about the verification of control parameter
calculations in CBTC as follows:

 \begin{itemize}
\item Modeling
\begin{itemize}
\item As discussed above, it is hard to build and verify the complete behavior space model of the control functions offline in advance. We propose that the model should focus on the ongoing static behavior of the system in the short future driven by the current control parameter.
\item For modeling the ongoing behavior of the running CBTC systems, as the system is composed by large number of components, e.g., one control system for each train running on track, the model should be a composed system naturally.
\item Data are transmitted along with communication between components. Thus, the modeling language needs to support the representation of the synchronization among components and the data transmission along with it.
\end{itemize}

\item Verification
\begin{itemize}
\item The verification problem will not try to prove whether the control functions are correct or not. The verification procedure will focus on giving answers of the correctness of current parameters.
\item As models are generated online, the verification procedure needs to be carried out online. As the model for the system will be updated quickly, it is necessary to give the verification result before the model is changed, which means the verification has to be time bounded and fast.
\item A set of parameters can basically induce a series of operation modes in the short future, which can consist several scenarios of the operation of the system. What need to be verified is the existence of certain scenarios in the behavior of the model, which is represented as the reachability of certain paths in the model.
\end{itemize}
\end{itemize}

Therefore, from both the point of views of modeling and
verification, in this paper we propose a new method to prove the
correctness of the parameter calculations online during the CBTC
system is in operation, which can result in an additional device
deployed on-site, monitoring and guaranteeing the correctness of parameters online. 

Based on this scheme, we present a formal model named as Hybrid
Automata with Readable Shared Variable to model the control
parameter calculations of the CBTC system, and a path-oriented
reachability verification technique to verify the reachability
property along with a path set in the model to achieve the goal of
fast online verification.  To demonstrate the feasibility of this
scheme,  the model for the control parameter calculations of the
CBTC system is given in the paper, and several case studies are
conducted on the model to illustrate the
performance of the fast online verification.\\

 \noop{\noindent {\bf
Contributions.} The contributions of this paper are summarized as
following:
\begin{itemize}
\item This paper investigates the requirements of verifying control parameter calculations in CPS systems using a running example of a CBTC system.
\item This paper presents a schema for online and fast verification of the ongoing model induced by the current control parameters to check the existence of the dangerous scenarios in the model.
\item A formal modeling language, Composed Linear Hybrid Automata with Readable Shared Variable, is given for modeling the control parameters in the CBTC system.
\item A path-oriented reachability analysis method is proposed for the verification of the existence of dangerous scenarios.
\item The model for the control parameter calculations of the CBTC system is given in the paper, and a series of case studies are conducted on the model to illustrate the performance of fast online verification.
\end {itemize}}

\noindent {\bf Structure of The Paper.} This paper is organized as
follows. In the next section, we give a brief description of the
running example of our study: Communication based train control
system and summarize the requirements of verifying the control
parameter calculations in a CBTC system. In Sec.3, we present our
modeling language for the CBTC system: Composed Hybrid System with
Readable Shared Variables, give the model we built for the CBTC
system and show how to verify the existence of given critical
scenarios on the system by the path-oriented reachability analysis
method.  Sec.4 verifies the existence of the dangerous scenarios in
the model we built for the CBTC system and demonstrates the process
ability of the path-oriented reachability method in online
verification of CBTC systems. Sec.5 summarizes the related works on
the verification of train control systems and proposes several
potential directions in the verification of CPS systems based on our
experience in verifying the CBTC system. Finally, the conclusion is
stated in Sec.6.

\section{Motivating Example: CBTC System}\label{sec:example}
\subsection{Communication Based Train Control System}
A train control system is the heart for the safe and efficient operation of train systems. There are many organizations and projects devoted to the research and development of the train control system with high dependability. Many standards are proposed to give detail and comprehensive rule sets and guidances for the operation of railway systems for inter-vehicle and vehicle to infrastructure cooperation, like European Train Control System(ETCS)\cite{etcs}, and Chinese Train Control System (CTCS)\cite{ctcs}.  According to different infrastructure utilities, data transmission methods and train control methodologies, ETCS/CTCS is divided up into several different equipmental and functional levels.

\begin{figure*}
\centering
\includegraphics[width=0.8\textwidth]{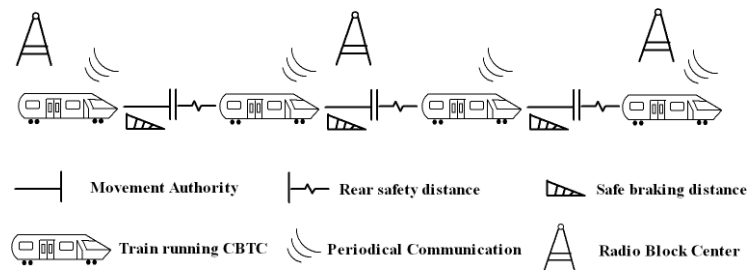}
\caption{Sample Scenario of A Running CBTC system}
\label{fig:cbtc}
\end{figure*}

Communication based train control system (CBTC), on the high level
of ETCS/CTCS, is believed to be the most advanced signaling
technique and the fundamental method underlying the latest high
speed railway systems. It uses data communication between trains and
various control facilities to guarantee the safety and efficiency of
train operation. It can be abstractly divided into two main parts:
ground systems and onboard systems. Ground systems can track the
runtime status of all the trains periodically. The radio block
center (RBC) will send the needed information, e.g., movement
authority, to the onboard systems on the train. Then the onboard
systems will compute the velocity curve autonomously by taking
account of the movement authority they received and the current
operation status of the train.  Ideally, the movement authority
basically indicates a End-of-Authority (EOA) point\cite{long,short}
which is with rear safe distance to the end of the train ahead.
During the train operation, it also needs to guarantee that there is
enough space for the train to completely stop by emergency braking
before touching the EOA point, which is named as ``Safe braking
distance''(SBD).  A simple illustration of the communication and
movement authority granting is shown in Fig.~\ref{fig:cbtc}.

Our running example is a typical CBTC system which is supposed to be
used in a Urban Railway System in China.  As the system is still
under designing and debugging, our team join into the project to
help verifying the correctness of the design of the ATP module of
the CBTC system. One of the most interested question which bothers
the engineers from the area of railway system is how to guarantee
the absence of certain dangerous scenarios, e.g., train collision.
We will use some of these scenarios to show the motivation of this
paper, and introduce our thoughts about the verification of control
parameter calculations in CBTC system.

In the design, all the trains need to communicate with RBC in 500
milliseconds period. RBC will grant the movement authority to each
train by telling them the position of the EOA points. After that,
the onboard computer will start to calculate the legal operation
speed by taking account of the current speed of train, the
limitation of the train and the track and so on. The train is free
to move under the generated operating speed before reaching the safe
braking distance point (SBD) which is with safe braking distance
away from the EOA point. Once a SBD point is reached, the train will
brake immediately to try to stop completely before move beyond the
EOA point. When the train has not receive any signal from RBC for 5
seconds, the automatic train protection (ATP) module of the CBTC
system will take over the control of the train operation and ask the
train to brake urgently as well. What the designers are worried
about is whether the train can stop safely under the control
parameters without beyond the movement authority and collide with
the train ahead under certain scenarios.


\subsection{Requirements For Verifying Control Parameter Calculations In The CBTC System}

For verifying the control parameter calculations in the CBTC system,
we need to build the formal model for the system and summarize the
characteristics of the CBTC system at first:

\begin {itemize}

\item {\bf Modeling}
\begin{itemize}

\item {\bf Static Model For Time-Bounded Behavior.}
As the behavior of the CBTC system is highly nondeterministic, e.g.,
the input of the control functions including lots of runtime dynamic
parameters, it is difficult to predict the complete behavior of the
system or even verify it. Thus, the modeling and verification effort
should focus on the ongoing behavior after synchronization or
receiving signals which is the deterministic part of the entire
nondeterministic behavior space.

\item {\bf Compositional Verification.} The running CBTC system has all the operating trains on track and RBCs included.
These components communicate with each other intensively, which is a concurrent system naturally. 
For each component, due to the dynamical behavior of system,
it should be a hybrid automata. Thus, the model for the system should be a composed hybrid system.

\item  {\bf Shared Label and Variable.} The modeling language has to support the representation of the
synchronization between components. Data are also transmitted during
the synchronization. This is natural, because a component running in
a system can not read the other components' running parameters at
anytime. 
We propose to use shared labels to represent the synchronization
among component automata, and communication with other components
are represented as shared variables in transition guards and reset
actions on shared labels.


\end{itemize}

\item{\bf Verification}
\begin{itemize}

\item {\bf Online and Fast Verification.} As the environment of the CBTC working in is changing quickly, 500 ms in the running example, if we cannot give answer to the verification questions in 500 ms,
the result will be meaningless. Thus, once a set of control
parameters is calculated, the verification module needs to give a
quick answer of whether this set of parameters will violate certain
properties, e.g., safety. Therefore, The verification should be
online and fast.

\item{\bf Control Parameters Driven Verification.} As the verification procedure needs to be online and fast,
it will not try to determine the correctness of the complex control
functions beneath the system, but only give a quick answer to the
correctness of the parameters generated.

\item {\bf Time-Bounded and Scenario-Based Verification.} According to the requirements of designers of the certain CBTC
system, the verification problem they concern most is checking
whether certain bad scenarios will happen in the control modes
induced by the current parameters. The scenario will be translated
as a sequence of control modes in the model, which constitutes a
path. Thus, what needs to be verified is the reachability of certain
property along with the path/scenario.
\end {itemize}
\end{itemize}

In summary, we think the modeling language for the control parameter
calculations in CBTC system is composed hybrid automata with support
of shared labels and shared variable reading.  The model of the
system should be a small static model induced by generated control
parameters. The verification procedure should be scenario based
path-oriented compositional reachability analysis.

\section{Verifying The Control Parameter Calculations in CTBC systems}


\subsection {Modeling of The CBTC System}

For a CBTC system in a train, the set of control parameters includes
the current velocity range, the target velocity range, the location
of the end of movement authority (EOA) point, the location of the
safe braking distance (SBD) point,  the position of the train itself
and so on.  Thus, this set of parameters shows a clear dynamic
behavior of the train along with time, which can be modeled as a
hybrid automaton (HA) naturally.

Now, let's raise the field of our view from a single train to a
series of trains running in a track. We will see that trains will
communicate with RBCs and other trains during operation
periodically. Data, e.g., the location of the train ahead, are
transmitted to each train along with the communication. Thus,  the
model for the complete system should be a composed hybrid automata.
Furthermore, in the composed system, component HAs synchronize with
each other using shared labels, and a component can only read the
value of the variable of other components on shared labels.

Based on the above discussion, we give the definition of the class
of HA we proposed for CBTC systems as following:

\begin{definition}\rm
A hybrid automaton with readable outer variables ($HA_{RV}$) is a
tuple $H=(X^l, X^s,\Sigma^l, \Sigma^s,V,V^0,E,\alpha,\beta,\gamma)$,
where
\begin{itemize}
\item[-]$X^l$ is a finite set of real-valued variables which belongs to $H$; $X^s$ is a finite set of real-valued variables which don't belong to $H$, but can be read by $H$ in certain position;  $X^l \cap X^s =\emptyset$
\item[-] $\Sigma^l$ is a finite set of local event labels which belongs to $H$ only;  $\Sigma^s$ is a finite set of event labels which belongs to several $HA_{RV}$; $\Sigma^l \cap \Sigma^s =\emptyset$
\item[-]$V$ is a finite
set of {\it locations}; $V^0\subseteq V$ is a set of \emph{initial
locations}.

\item[-]$E$ is a {\it transition relation} whose elements
are of the form $(v,\sigma,\phi,\psi,v')$, where $v,v'$ are in $V$,
$\sigma\in\Sigma^l\cup\Sigma^s$ is a label, $\phi$ is a set of
\emph{transition guards} of the form $f(\vec{y})\leq a$, and $\psi$
is a set of \emph{reset actions} of the form $x :=
 f(\vec{y})$, where $x\in X^l$, $a\in
\mathbb{R}$, and
\begin{itemize}
\item if $\sigma\in \Sigma^l$, $y\in X^l$;
\item if $\sigma \in \Sigma^s$, $y\in X^l\cup X^s $, if $y\in X^s$, we say $\sigma$ is $y$ related ;
\end{itemize}

\item[-]$\alpha$ is a labeling function which maps each location in
$V$ to a \emph{location invariant} which is a set of {\it variable
constraints} of the form $f(\vec{y})\leq a$ where $y\in X^l$, $a\in
\mathbb{R}$.
\item[-]$\beta$ is a labeling function which maps each location in
$V$ to a set of \emph{flow conditions} which are of the form
$\dot{x}= g(\vec{y})$ where $x\in X^l$. For any $v\in V$, for any
$x\in X^l$, there is one and only one flow condition $\dot{x} =
g(\vec{y})\in\beta(v)$, where $x,y\in X^l$.
\item[-]$\gamma$ is a labeling function which maps each location in
$V^0$ to a set of \emph{initial conditions} which are of the form
$x=a$ where $x\in X^l$ and $a\in \mathbb{R}$. For any $v\in V^0$,
for any $x\in X^l$, there is at most one initial condition
definition
$x = a\in\gamma(v)$. 
\end{itemize}
\end{definition}

If each $f(\vec{y})$ is a linear expression, and $g(y)=[a,b]$, where
$a,b\in \mathbb{R}$, we say this $HA_{RV}$ is a $LHA_{RV}$ (linear
hybrid automaton with readable outer variables).

For a group of $HA_{RV}$, their composition $CHA_{RV}$ is defined as
a product $HA_{RV}$ generated by synchronizing all the components
with respect to the shared labels.
\begin{definition}\rm
  Let $H_1=(X^l_1,X^s_1,\Sigma^l_1,\Sigma^s_1,V_1,V^0_1,E_1,\alpha_1,\beta_1,\gamma_1)$ and $H_2=(X^l_2,X^s_2,\Sigma^l_2,\Sigma^s_2,V_2\\,V^0_2,E_2,\alpha_2,\beta_2,\gamma_2)$ be
two $HA_{RV}$s, where $X^l_1\cap X^l_2=\emptyset$, $\Sigma^l_1\cap
\Sigma^l_2=\emptyset$. The {\it composition} of $H_1$ and $H_2$,
denoted as $H_1||H_2$, is a $HA_{RV}$
$N=(X^l,X^s,\Sigma^l,\Sigma^s,V,V^0,E,\alpha,\beta,\gamma)$ where
\begin{itemize}
\item $X^l=X^l_1\cup X^l_2$; $X^s=\{X^s_1\cup X^s_2\} \setminus \{ X^l_1\cup X^l_2\}$;

\item$\Sigma^l=\Sigma^l_1\cup\Sigma^l_2\cup \{\Sigma^s_1\cap\Sigma^s_2\}$;
$\Sigma^s=\Sigma^s_1\cup\Sigma^s_2$;

\item$V=V_1\times V_2$; $V^0=V^0_1\times V^0_2$;
\item $\alpha((v_1,v_2))=\alpha(v_1)\cup\alpha(v_2)$;
$\beta((v_1,v_2))=\beta(v_1)\cup\beta(v_2)$;
$\gamma((v_1,v_2))=\gamma(v_1)\cup\gamma(v_2)$;

\item $E$ is defined as follows:
\begin{itemize}
  \item[-] for $a \in \Sigma^s_1\cap\Sigma^s_2$, for every
  $(v_1,a,\phi_1,\psi_1,v'_1)$ in $E_1$ and $(v_2,a,\phi_2,\psi_2,v'_2)$ in
  $E_2$, $E$ contains
  $((v_1,v_2),a,\phi_1\cup\phi_2,\psi_1\cup\psi_2,(v'_1,v'_2))$;
  \item[-] for $a \in \Sigma^l_1\cup\{\Sigma^s_1\setminus\Sigma^s_2\}$, for every
  $(v,a,\phi,\psi,v')$ in $E_1$ and every $t$ in
  $V_2$, $E$ contains $((v,t),a,\phi,\psi,(v',t))$;
  \item[-] for $a \in \Sigma^l_2\cup\{\Sigma^s_2\setminus\Sigma^s_1\}$, for every
  $(v,a,\phi,\psi,v')$ in $E_2$ and every $t$ in
  $V_1$, $E$ contains $((t,v),a,\phi,\psi,(t,v'))$.
\end{itemize}
\end{itemize}
For all $m>2$, the {\it composition} of $HA_{RV}$ $H_1,H_2,\dots,
H_m$, denoted as $H_1||H_2||\dots ||H_m$, is a $HA_{RV}$ which is
defined recursively as $H_1||H_2||\dots ||H_m=H_1||H'$ where
$H'=H_2||H_3||\dots ||H_{m}$.
\end{definition}

\begin{figure*}[!h]
\centering
\includegraphics[width=1.0\textwidth]{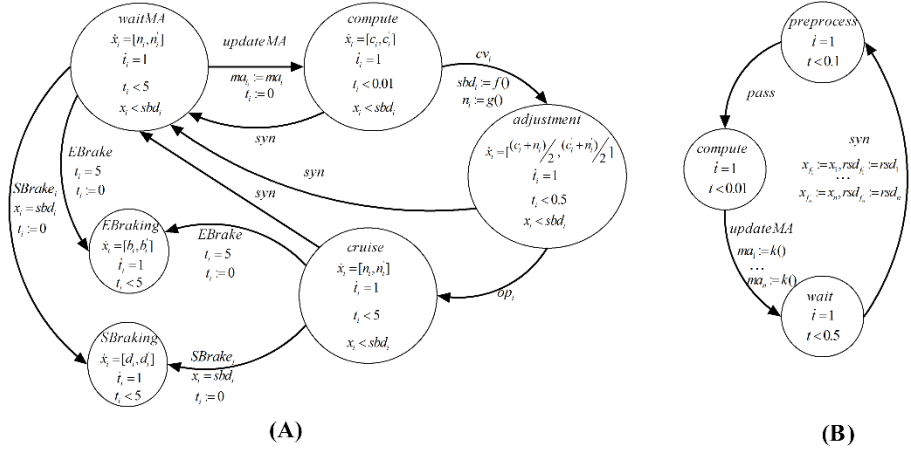}
\caption{Hybrid Automata For $Train_i$ Using CBTC and RBC Center
$RBC_j$} \label{fig:cps_all}
\end{figure*}

Using the formal language defined above, 
we build a set of models of the system which includes nonlinear
control functions as shown below. These models consist of two main
parts:
\begin{itemize}

\item $n$ trains running on the track, the automaton for each train is shown in Fig.\ref{fig:cps_all}.A.
\item $m$ RBC centers, 
the automaton for each RBC is shown in Fig.\ref{fig:cps_all}.B.

\end{itemize}

From Fig.\ref{fig:cps_all}, we can see the behavior of the system
consists of the following aspects:
\begin{itemize}

\item Trains and RBCs communicate by two labels $updateMA$ and
$syn$.

\item After the global synchronization $syn$, an RBC will get the running parameters from
the related train. Then the RBC will perform preprocess job before
it starts to compute and assign the latest MA to the related trains.

\item After preprocessing, RBC will compute the new MA
for the related trains using complex function $k()$ and send them to
the related trains by shared label $updateMA$.

\item When received the new MA, each train will compute the local velocity
and SBD using control function $f()$ and $g()$, and it will start to
adjust the running velocity from current value $[c_i,c'_i]$ to the
latest value range $[n_i,n'_i]$.

\item During the adjustment period, we abstract the velocity to the mean of the old and new value of the velocity range.

\item After the train $Train_i$ is running under the new velocity range, it will keep checking the current position to make sure it has not move beyond the safe braking point.

\item Once the safe braking point is touched, the train will start to brake normally to try to stop completely before touching the end of the movement authority.

\item And if the train has safely operated for 5 seconds without receiving any communication signal, the train will assume the communication channel is broken and an emergency braking will be executed immediately.

\item Once a train starts the procedure of braking, it must stop completely in less than 5 seconds.

\end{itemize}

Considering a system with dozens of subsystems, e.g., trains and
RBCs, and with complex nonlinear functions $f(),g(),k()$ included,
it will be very difficult to verify properties on the model, as
widely reported in
literature~\cite{toolsurvey}. 
Furthermore, many parameters used in functions $f(),g()$ and $k()$
are collected and generated online nondeterministically, e.g.,
temporary speed limitation, wind speed, mass of the train and etc,
even there is a method to verify the complex nonlinear function, as
these critical parameters cannot be predicted ahead precisely, the
offline verification of the system is still very difficult.

\noop{Using the formal model defined above, the model we built for
the CBTC system running on train $Train_i$ is shown below in
Fig.\ref{fig:cps}. This hybrid automaton models the local behavior
of  $Train_i$ according to the control parameters generated after
the global synchronization. The complete system is composed by $n$
models for $Train_1$ to $Train_n$. Since the control function for
computing the movement authority in RBC will not be included in the
model, then it is not necessary to build a model for RBC in the
system. We can express the communication and data transmission of
RBC by shared labels between trains along with shared variable
reading on the shared labels, for example on shared label $syn$, the
movement authority of $Train_i$ is reset by minus the location of
the train ahead by the rear safe distance: $ma_i:=x_{i-1}-rsd_i$.}

\subsection{Scenario Based Verification and Path-oriented Reachability Analysis}

For the verification of the control parameter calculations in CBTC
systems, one of the problems which the designers concern most is
when $Train_i$ starts to brake, whether it can stop completely
before passing the EOA point or even collide with the ahead train
under the generated parameters. This problem indicates an execution
scenario of the behavior of each train in the system, from location
$compute$ to $Ebraking$, and a target property to verify: the
physical
position of $Train_{i}$ equals with the ahead one $Train_{i-1}$.\\

\noindent{\bf{Scenario-Based Automata.}} As discussed in the last
section, the verification of the given scenario-based property on
the models given in Fig.\ref{fig:cps_all} is very difficult. On the
other hand, the control parameters generated by the control
functions can induce a static control model of the behavior of the
CBTC system in the short future before the generation of the next
set of parameters. 

Based on this idea, we simplify the models given in the last
section. As the control parameters are already calculated and saved,
e.g. $ma_{c_i}$ and $sbd_{c_i}$ for the movement authority and safe
braking distance of $Train_i$, the control functions can be
dismissed in the new scenario-based model. The component RBCs can
also be dismissed from the system, because the scenario-based
automaton stands for the behavior of the system after the latest MA
is already granted and before the next communication, during that
period the train $Train_i$ doesn't have to communicate with any RBC.
As a result, we build the scenario-based static running automata for
$Train_i$ to a $LHA_{RV}$ as below in Fig.\ref{fig:cps_sce}.

\begin{figure*}[!h]
\centering
\includegraphics[width=0.9\textwidth]{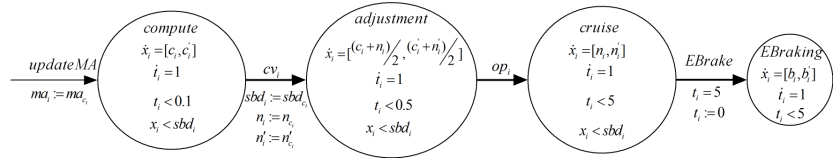}
\caption{Scenario Based $LHA_{RV}$ For $Train_i$ Using CBTC}
\label{fig:cps_sce}
\end{figure*}

This scenario on a single $LHA_{RV}$ $Train_i$ is presented as an evolution of the system from locations to locations, e.g., 
 $\langle comp\rangle{\underset{cv_i}\longrightarrow}\langle adjust\rangle{\underset{op_i}\longrightarrow}\langle cruise\rangle{\underset{EBrake}\longrightarrow}\langle EBrake\rangle$ in the automaton $Train_i$. Using the same notion given in \cite{sttt}, we name such a sequence of locations as a {\it path}. By assigning each location with an nonnegative real number, we can get a {\it timed sequence} in the form of
$\left\langle\begin{array}{l}
comp\\\delta_0\end{array}\right\rangle{\underset{cv_i}\longrightarrow}\left\langle\begin{array}{l}
adjust\\\delta_1\end{array}\right\rangle{\underset{op_i}\longrightarrow}\left\langle\begin{array}{l}
cruise\\\delta_2\end{array}\right\rangle{\underset{EBrake}\longrightarrow}\left\langle\begin{array}{l}
EBrake\\\delta_3\end{array}\right\rangle$. This timed sequence
represents a {\it behavior} of the model such that the system starts
at location $compute$, stays there for $\delta_0$ time units, then
jumps to location $adjust$ by transition $cv_i$ and stays at
$adjust$ for $\delta_1$ time units, and so on.

\noop{More generally, the behavior of an $LHA_{RV}$ can be described
informally as follows. The automaton starts at one of the initial
locations $v_0$ with all the variables $\vec{x}$ initialized to
their initial values $\gamma_0(\vec x)$. As time progresses, the
values of all variables change continuously according to the flow
condition associated with the current location. At any time, the
system can change its current location from $v_i$ to $v_{i+1}$
provided that there is a transition
$(v_{i},\sigma,\phi,\psi,v_{i+1})$ from $v_{i}$ to $v_{i+1}$ whose
all transition guards in $\phi$ are satisfied by the current value
$\zeta_i(\vec x)$ of all the variables $\vec{x}$. If the location is
changed by a transition $(v_i,\sigma,\phi,\psi,v_{i+1})$, some
variables are reset to the new values according to the reset actions
in $\psi$, so that all the variables get a new set of valuations
$\gamma_{i+1}(\vec x)$. Transitions are assumed to be instantaneous.
Finally the system will stop in location $v_n$ with the valuation
set $\zeta_n({\vec x})$ of all the variables.}

Let $N=H_1||H_2||\dots||H_m$ be a $CLHA_{RV}$ where 
$H_i=(X^l_i,X^s_i,\Sigma^l_i,\Sigma^s_i,V_i,V^0_i,E_i,\alpha_i,\beta_i,\gamma_i)$ $(1\leq
i\leq m)$ is an $LHA_{RV}$ 
and $\rho$ be a path in $N$ of the form $\rho=\langle
v_0\rangle\overset{(\phi_0,\psi_0)}{\underset{\sigma_0}\longrightarrow}
\langle
v_1\rangle\overset{(\phi_1,\psi_1)}{\underset{\sigma_1}\longrightarrow}
\dots\overset{(\phi_{n-1},\psi_{n-1})}{\underset{\sigma_{n-1}}\longrightarrow}\langle
v_{n}\rangle$. It follows that $v_i=(v_{i1},v_{i2},\dots,v_{im})\
(0\leq i\leq n)$ where $v_{ik}\in V_k\ (1\leq k\leq m)$. For any $k$
$(1\leq k\leq m)$, we construct the sequence $\rho_k$ from $\rho$ as
follows: replace any $v_i$ with $v_{ik}$ $(0\leq i\leq n)$, and for
any
$\overset{(\phi_{i-1},\psi_{i-1})}{\underset{\sigma_{i-1}}\longrightarrow}
\langle v_{ik}\rangle\ (1\leq i\leq n)$, if
$(v_{i-1k},\sigma_{i-1},\phi,\psi,v_{ik})\in E_k$, then replace it
with $\overset{(\phi,\psi)}{\underset{\sigma_{i-1}}\longrightarrow}
\langle v_{ik}\rangle$, otherwise remove it. It follows that
$\rho_k$ is a path in $H_k$. We say that $\rho_k$ is the {\it
projection} of $\rho$ on $H_k$. Intuitively, $\rho_k$ is the
execution trace of $N$ on $H_k$ when $N$ runs along $\rho$. Thus,
the complete scenario is a path set for the system, consisting of
one path for each component.\\

\noindent{\bf{Reachability Specification.}} Now, let us look at the
reachability specification: During braking, $Train_{i}$ collide with
the ahead train $Train_{i-1}$, which means the position of
$Train_{i}$ is the same with the ahead one $Train_{i-1}$ in the
location $EBraking$. This property can be formally translated as
$Train_i.x=Train_{i-1}.x$ in location $EBraking$.

For an $LHA_{RV}$ $H=(X^l,X^s,\Sigma^l,\Sigma^s,V,V^0,E,\alpha,\beta,\gamma)$, a {\it
reachability specification}, denoted as ${\cal R}(v,\varphi)$,
consists of a location $v$ in $H$ and a set $\varphi$ of variable
constraints of the form $a\leq c_0x_0+c_1x_1+\dots+c_lx_l\leq b$
where $x_i\in X^l\cup X^s$ for any $i\ (0\leq i\leq l)$, $a,b$ and $c_i\
(0\leq i\leq l)$ are real numbers.

\begin{definition}\rm
Let $H=(X,\Sigma,V,V^0,E,\alpha,\beta,\gamma)$ be an $LHA_{RV}$, and
${\cal R}(v,\varphi)$ be a reachability specification. A behavior of
$H$ of the form $\left\langle\begin{array}{l}
v_0\\\delta_0\end{array}\right\rangle\overset{(\phi_0,\psi_0)}{\underset{\sigma_0}\longrightarrow}
\left\langle\begin{array}{l}
v_1\\\delta_1\end{array}\right\rangle\overset{(\phi_1,\psi_1)}{\underset{\sigma_1}\longrightarrow}
\dots\overset{(\phi_{n-1},\psi_{n-1})}{\underset{\sigma_{n-1}}\longrightarrow}\left\langle\begin{array}{l}
v_n\\\delta_n\end{array}\right\rangle$ {\it satisfies} ${\cal
R}(v,\varphi)$ iff $v_n=v$ and each variable constraint in $\varphi$
is satisfied when the automaton has stayed in $v_n$ for delay
$\delta_n$, i.e. for each variable constraint $a\leq
c_0x_0+c_1x_1+\dots+c_lx_l\leq b$ in $\varphi$, $a\leq
c_0\zeta_n(x_0)+c_1\zeta_n(x_1)+\dots+c_m\zeta_n(x_l)\leq b$
 where $\zeta_n(x_k)\ (0\leq k\leq l)$ represents the value of
 $x_k$ when the automaton has stayed at $v_n$ for the delay
 $\delta_n$. $H$ {\it satisfies} ${\cal R}(v,\varphi)$ iff there is a behavior of $H$ which satisfies ${\cal
R}(v,\varphi)$.
\end{definition}


\begin{definition} \rm Let $N=H_1||H_2||\dots||H_m$ be a $CLHA_{RV}$,
$P=\{\rho_1,\rho_2,\dots,\rho_m\}$ be a path set, where $\rho_i$ is
a finite path in $H_i\ (1\leq i\leq m)$, and ${\cal R}(v,\varphi)$
be a reachability specification. $P$ {\it satisfies} ${\cal
R}(v,\varphi)$ if and only if there is a path $\rho$ of $N$ that the
projection of $\rho$ on $H_i$ is $\rho_i\ (1\leq i\leq m)$, and
there is a behavior of $N$ which satisfies ${\cal R}(v,\varphi)$.\\

%
\end{definition}

\noindent{\bf{Path-Oriented Reachability Analysis.}} In this
paragraph, we will show how to verify the reachability specification
along with a path set in a $CLHA_{RV}$ system using linear
programming efficiently.

Generally speaking, the model checking problem for hybrid systems is
very difficult. Even for a single LHA, the reachability analysis
problem is undecidable~\cite{tth,rcnt,tpap}. The performance of
existing techniques for compositional analysis of LHA systems is
even worse.  The state-of-the-art tool HYTECH~\cite{hytech} and its
improvement PHAVer~\cite{phaver} need to compute the composition of
the whole system into a unique global automaton then use expensive
polyhedra computation for reachability analysis, which will suffer
the problem of state explosion and greatly restrict the solvable
problem size.

To overcome this drawback, in study\cite{sttt} we presented an efficient approach for the path-oriented reachability analysis of LHA compositions. This technique checks a group of paths at a time, one path for each LHA,  all of the paths are transformed into a group of
linear constraints automatically. Then, a few constraints about the
system integration according to the synchronization events in each
path will be added to ensure that the components cooperate
correctly. It follows that the reachability problem along those
specific paths can be reduced to a linear program.
Using this method both the path
length and the number of participant automata checked can be scaled up
greatly to satisfy practical requirements. This approach of symbolic
execution of paths can be used by design engineers to check critical
paths, and thereby increases the faith in the system correctness.
This path-oriented technique can be easily scaled to use in $CLHA_{RV}$ systems. We will use a
simple example to illustrate our idea below.

\begin{figure*}[ht]
\centering
\includegraphics[width=1.0\textwidth]{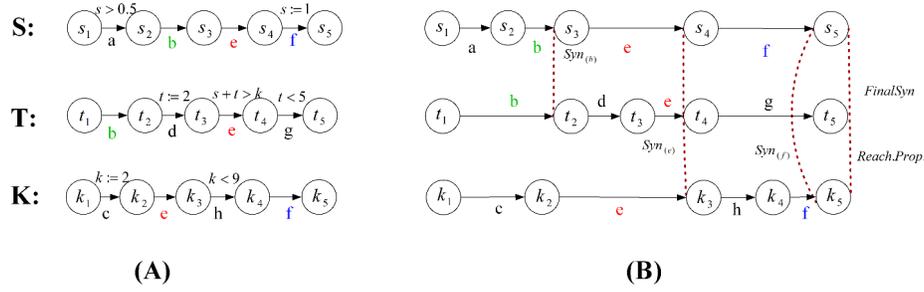}
\caption{Sample Automata And The Path-oriented Reachability Encoding} \label{vs}

\end{figure*}

For example, Fig.\ref{vs}(A) gives a simple system consisting of
three subsystems: $S$, $T$, and $K$ which synchronize with each
other by shared labels $b$, $e$, and $f$. Each system has one
variable, $s$ for $S$, $t$ for $T$, $k$ for $K$. The flow conditions
for each variable are unified as $\dot{\vec x}\in [0.9, 1.1]$ in all
the locations.
The values of data are transmitted along with some of the shared labels, for example in label $e$ of $T$, the transition guard is $s+t>k$. The reachability specification is whether the property $s+2t-3k=0$ can be satisfied at the global location $(s_5,t_5,k_5)$. 

In our path-oriented approach,
for each of these three paths we generate a group of linear
constraints that represents all the timed runs corresponding to  the
path.  Take  the path $\langle t_1\rangle \rightarrow
\langle t_2\rangle \rightarrow \langle t_3\rangle \rightarrow
\langle t_4\rangle \rightarrow \langle t_5\rangle $ of the system
$T$ for example:

\begin{itemize}
\item Use $\left\langle\begin{array}{l}
t_i\\\delta^t_i\end{array}\right\rangle$ to indicate that the system
has stayed in location $t_i$ for time delay $\delta_i$ (nonnegative
variable). The behavior of the system is
represented by
$\left\langle\begin{array}{l} t_1\\\delta^t_1\end{array}\right\rangle\rightarrow
\left\langle\begin{array}{l}
t_2\\\delta^t_2\end{array}\right\rangle\rightarrow
\left\langle\begin{array}{l}
t_3\\\delta^t_3\end{array}\right\rangle\rightarrow
\left\langle\begin{array}{l}
t_4\\\delta^t_4\end{array}\right\rangle\rightarrow
\left\langle\begin{array}{l} t_5\\\delta^t_5\end{array}\right\rangle$
\noindent where $\delta^t_1,\delta^t_2,\delta^t_3,\delta^t_4,\delta^t_5$ must
satisfy all the time constraints enforced by the system, which forms
a group of linear constraints.
\begin{itemize}
\item For each location $t_i$, two variables $\gamma_i(t)$ and $\zeta_i(t)$ are generated to represent the valuation of $t$ when entering $t_i$ and leaving $t_i$ after stay there by $\delta_i$ time units.
\item Take the location $t_3$ for example, according to the flow condition, $ 1.1\delta^t_3+ \gamma_3(t)\geq\zeta_3(t)\geq 0.9\delta^t_3+ \gamma_3(t)$.
\item For the transition guard $t<5$ on the local transition $g$, we have $\zeta_4(t)<5$.
\item For the reset action $t=2$ on the local transition $d$, we have $\gamma_3(t)=2$.

\end{itemize}

\item Synchronization constraints will be
added to ensure that these three components cooperate accurately
according to the synchronization events, which are illustrated by
the dashed lines and $SYN_{(event)}$ in Fig.\ref{vs}(B).
\begin{itemize}
\item For the event $b$ shared by $S$ and $T$, we have $\delta^t_1=\delta^s_1+\delta^s_2$.
\item For the transition constraints including outer variable reading, e.g., $s+t>k$ in $e$, we have $\zeta_3(s)+\zeta_3(t)>\zeta_2(k)$.
\item All the components have spent exact the same time, e.g., for $S$ and $T$, we have $\delta^s_1+\delta^s_2+\delta^s_3+\delta^s_4+\delta^s_5=\delta^t_1+\delta^t_2+\delta^t_3+\delta^t_4+\delta^t_5$.
\item For reachability specification $s+2t-3k=0$,  we get $\zeta_5(s)+2\zeta_5(t)-3\zeta_5(k)=0$.
\end{itemize}

\end{itemize}

Above all, the path-oriented reachability analysis problem is
transformed to a feasibility problem of a set of linear constraints.
It is well-known that the feasibility problem of linear constraints
can be solved by linear programming (LP) technique efficiently. 
Utilizing LP solver, we can develop an efficient tool for
path-oriented reachability analysis of $CLHA_{RV}$ where the length
of the path, the size of each $LHA_{RV}$, and the number of
components are all close to the practical problem scales. Thus, we
can gain the objective of fast verification of the existence of
certain scenarios in the model of control parameter calculations in
CBTC systems.

\section{Experimental Evaluation}
\label{sec:case}

To demonstrate the modeling and verification techniques for control parameter calculations in the CBTC system proposed in this paper and show the ability of fast verification of the path-oriented reachability method, we verify the train collision scenario given in the last section using the model built in Sec.3.

The scenario we selected to verify is if the communication channel
fails during train operation, whether all the train can stop safely
without collide with each other, and the corresponding
scenario-based automata we built for each train is shown in
Fig.\ref{fig:cps_sce}. The model represents the path: $\langle
compute\rangle{\underset{cv_i}\longrightarrow}$ $\langle
adjustment\rangle{\underset{op_i}\longrightarrow}$ $\langle
cruise\rangle{\underset{EBrake}\longrightarrow}\langle
EBraking\rangle$ for each train $Train_i$ and the reachability
specification is the positions of two nearby trains are equal with
each other, for example  $train_1.x=train_2.x$. Since the system is
still under simulation and debugging, we use a group of traditional
running values for the parameters in the model from our colleagues
in the railway area.

\noop{ The experiments are conducted in an ongoing version of
BACH\cite{blfmcad,bldate}, which can be downloaded from
\url{http://seg.nju.edu.cn/BACH/}. BACH is a toolset for building
LHA models and verifying the bounded reachability property of LHA
systems. BACH includes a graphical LHA editor which allows users to
construct, edit, and perform syntax analysis of LHA interactively, a
path-oriented reachability checker to check whether the reachability
specification is satisfied along with the given path set and a
bounded reachability checker to verify the specification of the
system in the given bound limit.}

The experiments are conducted in an ongoing version of
BACH\cite{blfmcad,bldate}, which is a toolset for building LHA
models and verifying the bounded reachability property of LHA
systems, and can be downloaded from
\url{http://seg.nju.edu.cn/BACH/}. On a DELL workstation (Intel
Core2 Quad CPU $2.4\mathrm{GHz}, 4\mathrm{GB}$ RAM), we evaluate the
potential of the path-oriented reachability analysis method
presented in this paper using the CBTC model shown in
Fig.\ref{fig:cps_sce}.

The experiment data is shown in Table.\ref{data}. The largest
problem BACH can solve in 500 ms consists of 16 trains which is a
very complex system and enough for a running urban railway system.
According to the consultation to the engineers in the urban railway
company, it is expected that the number of trains under operation on
a normal track is around 15 to 20. Thus, the technique presented in
this paper is applicable to be used in daily operation. The
parameter we used in the model is proved to be safe by verification,
which means certain path-oriented reachability specifications are
not satisfied. Meanwhile, the runtime memory overhead of the
computation, which is not listed in the table, is very small.

The data in Table.\ref{data} gives a clear demonstration of the
process ability of fast verification of the bad scenario in the
model for control parameter calculations. It also strengthens our
belief that this technique can be used online during system is in
operation to guarantee the correctness of the important control
parameters. As the linear programming solver underlying BACH is a
free collection of Java classes for research~\cite{oro}, we believe
if the linear programming package is replaced by an advanced
commercial one, the performance will be even better.

\begin {table}[!h]
\centering \caption{Experimental Data on the CBTC System}\scriptsize
\label{data}
\begin{tabular}[h]{|c|c|c|c|}\hline

&$Train_1$&\multicolumn{2}{|c|}{$\langle
compute\rangle{\underset{cv_1}\longrightarrow}\langle
adjustment\rangle{\underset{op_1}\longrightarrow}\langle
cruise\rangle{\underset{EBrak}\longrightarrow}\langle
EBraking\rangle$}\\
\cline{2-4} &$Train_2$&\multicolumn{2}{|c|}{$\langle
compute\rangle{\underset{cv_2}\longrightarrow}\langle
adjustment\rangle{\underset{op_2}\longrightarrow}\langle
cruise\rangle{\underset{EBrak}\longrightarrow}\langle
EBraking\rangle$}\\
\cline{2-4}
Path&\ldots&\multicolumn{2}{|c|}{$\ldots$}\\
\cline{2-4} &$Train_{n-1}$&\multicolumn{2}{|c|}{$\langle
compute\rangle{\underset{cv_{n-1}}\longrightarrow}\langle
adjustment\rangle{\underset{op_{n-1}}\longrightarrow}\langle
cruise\rangle{\underset{EBrak}\longrightarrow}\langle
EBraking\rangle$}\\
\cline{2-4} &$Train_n$&\multicolumn{2}{|c|}{$\langle
compute\rangle{\underset{cv_n}\longrightarrow}\langle
adjustment\rangle{\underset{op_n}\longrightarrow}\langle
cruise\rangle{\underset{EBrak}\longrightarrow}\langle
EBraking\rangle$ }\\\hline

$n$&Constraint&Variable&Time\\
\hline

8&1208&96&0.175s\\\hline
10&1550&120&0.23s\\\hline
12&1908&144&0.328s\\\hline
14&2282&168&0.404s\\\hline
16&2672&192&0.469s\\\hline

\end{tabular}
\end {table}

\noop{
\begin {table}[!h]
\centering \caption{Experimental Data on the CBTC System}\scriptsize
\label{data}

\begin{tabular}[h]{|c|c||c|c|c|c|}\hline

\multicolumn{2}{|c||}{Path}&$n$&Constraint&Variable&Time\\
\hline

 $Train_1$&{$\langle
compute\rangle{\underset{cv_1}\longrightarrow}\langle
adjustment\rangle{\underset{op_1}\longrightarrow}\langle
cruise\rangle{\underset{EBrak}\longrightarrow}\langle
EBraking\rangle$}&8&1208&96&0.175s\\\hline

$Train_2$&{$\langle
compute\rangle{\underset{cv_2}\longrightarrow}\langle
adjustment\rangle{\underset{op_2}\longrightarrow}\langle
cruise\rangle{\underset{EBrak}\longrightarrow}\langle
EBraking\rangle$}&10&1550&120&0.23s\\\hline

\ldots&{$\ldots$}&12&1908&144&0.328s\\\hline

$Train_{n-1}$&{$\langle
compute\rangle{\underset{cv_{n-1}}\longrightarrow}\langle
adjustment\rangle{\underset{op_{n-1}}\longrightarrow}\langle
cruise\rangle{\underset{EBrak}\longrightarrow}\langle
EBraking\rangle$}&14&2282&168&0.404s\\\hline

$Train_n$&{$\langle
compute\rangle{\underset{cv_n}\longrightarrow}\langle
adjustment\rangle{\underset{op_n}\longrightarrow}\langle
cruise\rangle{\underset{EBrak}\longrightarrow}\langle
EBraking\rangle$ }&16&2672&192&0.469s\\\hline

\end{tabular}

\end {table}
}

\section{Related Work and Further Discussion}

\subsection{Related Work}

The verification of the train control system has been intensively
studied. Study\cite{icse} gives a method to generate the high level
requirements from a subset of the specification of a ETCS\cite{etcs}
system, and use method in\cite{cav} to verify the consistency
between requirements. These two works belong to the category of
requirement engineering, which don't touch real time behaviors of
the system.

Study\cite{statemate} models the communication in train control
systems with Live Sequence Chart (LSC), then validate the LSC by
model checking and testing. Study\cite{formats} models the behavior
of train control systems by timed state transition systems and
verify the given property by bounded model checking and
compositional reasoning. These studies all give high level models
for behaviors of the system without considering the dynamic behavior
of the movement of train.

Study\cite{short} models a fully parametric ETCS system using
differential dynamic logic and verify the system by logical
deductive verification. Study\cite{long} builds different complex
models for different layers of a ETCS system and verify these models
using layer-specific technologies.  These works build static model
for the ETCS system without considering the system as a dynamic
system which works in open environment. Thus, they only include
rather limit parameters used in the control functions in the model.

\subsection{Verification of CPS Systems}

The new CPS computing paradigm brings new challenges and
requirements to the research community, like how to guarantee the
qualities of service, how to generate the formal models for the
system and so on, which are proposed and summarized in many studies
like \cite{lee2,ebar}. The CBTC system is a typical CPS system which
combines communication, computation and control tightly. From the
experience we learned during verify the CBTC system, we think {\bf
Control Parameter Calculations Verification} could be an emerging
topic in the verification of CPS systems. Furthermore, we summarize
following subtopics we think is worth studying and paying attention
to:

\begin{itemize}
\item {\bf Modeling Language.} CPS systems are running under dynamic environments. They receive signals from each other and the environment in a unpredictable way. How can the nondeterminism be modeled and verified? For CBTC systems, we choose to use linear hybrid automata as the modeling language and focus on the modeling of the ongoing static behavior of the system once the control parameters are generated. How about for general CPS system, do we need to introduce an new language?

\item {\bf Time Bounded Verification.} Compared with classical verification which try to prove the correctness of the complete behavior of the system, the verification of CPS system focuses more on the correctness of the behavior in given time bound, e.g., will the train collide with the ahead one in 500 milliseconds in this paper. This is a new direction of Bounded Model Checking\cite{bmc}, where the term ``bound'' means time, rather than ``steps'' used in classical Bounded Model Checking.

\item {\bf Online and Fast Verification.} As the control parameters of CPS system are changing quickly, the verification module needs to give a quick answer of the correctness of the new generated set of parameters. We think it is necessary to investigate how to build fast and low-overhead online verification techniques for CPS systems.

\end{itemize}

\section{Conclusion}
\label{sect:conclusion}

In this paper, we introduce our experience in modeling and verifying
the control parameter calculations in a CBTC system which is a
typical CPS system.  Based on our study of this system, we propose
our ideas of the requirements for modeling and verifying control
parameters in a CBTC system. For modeling language, we think it
should be a composed hybrid system with support of component
communication and data transmission. For verification technique, we
insist the verification for CBTC systems should be online and fast
verification of the ongoing behavior in the short future, and the
problem needed to be verified is the existence of certain dangerous
scenarios.

To demonstrate our ideas, we introduce a notion Composed Linear
Hybrid Automata with Readable Shared Variables to model the behavior
of the CBTC system induced by the control parameters. We also
present a path-oriented reachability analysis method to achieve the
objective of the online scenario-based verification.  The experiment
results support our belief a lot by showing the great process
ability of fast solving of a system consists of 16 trains in less
than 500 milliseconds which is the period of parameter generation in
the CBTC system.

Currently, with the help of our colleagues from railway areas, we
are trying to implement this technique into a standalone device
which can be integrated and deployed into the onboard ATP module as
a part of the CBTC system to check the correctness of the velocity
range given by ATP. Safety critical scenarios can be enumerated by
CBTC engineers ahead, the model pattern corresponding to these
scenarios can be designed in advance also. Then the device can catch
the latest generated parameter set, build the related models using
the pattern and verify them online. It is supposed to work as a
runtime monitor/checker on the train under experimentation to
guarantee the safety of the control parameters before the parameters
are utilized.

\end{document}